\documentstyle[preprint,aps,epsfig]{revtex}
\tightenlines
\newcommand{\bea}{\begin{eqnarray}}
\newcommand{\eea}{\end{eqnarray}}
\newcommand{\beq}{\begin{equation}}
\newcommand{\eeq}{\end{equation}}
\newcommand{\bay}{\begin{array}}
\newcommand{\eay}{\end{array}}
\begin{document}
\preprint{\parbox{6cm}{\flushright CLNS 98/1569}}
\title{Hybrid Baryons in Large-$N_c$ QCD}
\author{Chi-Keung Chow$^a$, Dan Pirjol$^b$\footnote{Address after Sept.~1$^{st}$:
Floyd R. Newman Laboratory of Nuclear
Studies, Cornell University, Ithaca, New York 14853} 
and Tung-Mow Yan$^a$}
\address{$^a$Floyd R. Newman Laboratory of Nuclear
Studies, Cornell University, Ithaca, New York 14853\\
$^b$Department of Physics,
Technion - Israel Institute of Technology, 32000 Haifa, Israel}
\date{\today}
\maketitle

\begin{abstract}
We study the properties and couplings of hybrid baryons in the
large-$N_c$ expansion. These are color-neutral baryon states which 
contain in addition to $N_c$ quarks also one constituent gluon. 
Hybrid baryons with both symmetric and mixed symmetric orbital wave
functions are considered. 
We introduce a Hartree description for these states, similar to the
one used by Witten for ordinary baryons. It is shown that the Hartree
equations for $N_c\,\,(N_c-1)$ quarks for symmetric (mixed symmetric) states 
in these states coincide with those in ordinary baryons in the large-$N_c$
limit. The energy due to the gluon field is of order $\Lambda_{QCD}$.
Under the assumption of color confinement, our results
prove the existence of hybrid baryons made up of
heavy quarks in the large $N_c$ limit and provides a justification
for the constituent gluon picture of these states. The couplings
of the hybrid baryons to mesons of arbitrary spin are computed in 
the quark model. Using constraints from the large $N_c$ scaling laws for
the meson-baryon scattering amplitudes, we write down consistency conditions 
for the meson couplings of the hybrid baryons. These consistency conditions
are solved explicitly with results in agreement with those in the quark
model for the respective couplings.
\end{abstract}
\pacs{pacs1,pacs2,pacs3}

\narrowtext

\section{Introduction}

It has been suggested for some time (for a review see \cite{3}) that a new 
class of hadron states could exist beyond the framework of the quark model. 
These are the so-called hybrid or hermaphrodite hadrons and contain, in 
addition to constituent quarks, also constituent gluons. From an experimental 
point of view it is easiest to distinguish a hybrid meson from an usual 
one. This is made possible by the fact that the former can have exotic 
quantum numbers, which are forbidden for states containing only constituent 
quarks (e.g. $J^{PC}=1^{-+}, 0^{--}$).

Perhaps for this reason, most of the theoretical effort has been devoted to the
study of the hybrid mesons. Their properties 
have been investigated extensively using QCD-inspired methods such as the bag model 
\cite{1,2}, the flux tube model \cite{IsPa,Page} and lattice QCD. Recently, some evidence
has been reported by the E852 Collaboration for the existence of a $J^{PC}=1^{-+}$
resonance \cite{E852}.

In addition to hybrid mesons, hybrid baryons are expected to exist too. These are
color-neutral states made up of $N_c$ quarks, each transforming under the fundamental
representation of $SU(N_c)_{color}$ plus one constituent gluon transforming in the
adjoint representation. The nontrivial color correlations between quarks and gluons 
constrain, via the Pauli principle, the spin-flavor structure of these states
to be different from that of pure $(q^{N_c})_{singlet}$ states. 

The lowest-lying hybrid baryons are expected to form a positive parity {\bf 70} 
multiplet of SU(6) lying around 1.5 GeV \cite{1,2} and have the constituent gluon
field in a TE mode with $J^P=1^+$. It is tempting to identify
the Roper resonance $N(1440)$ with such a hybrid baryon. Although this identification
is not yet completely ruled out, (see \cite{3} for a discussion of the various 
possibilities) it appears to conflict with existing photoproduction data.
An analog of the Moorhouse selection rule derived in \cite{1} forbids the 
photoproduction of a hybrid $J^P=1/2^+$ state off protons, whereas experimentally
the $N(1440)$ is photoproduced copiously.
A similar reasoning can be used to argue that the nearby $I,J^P=1/2,3/2^+$ state
$N(1720)$ cannot be predominantly hybrid. This leaves as the only hybrid candidate 
the $I,J^P=1/2,1/2^+$ state $N(1710)$.

The apparent absence of the hybrid baryons in this region is somewhat puzzling 
and raises the question of the very existence and stability of these states.
It is therefore of some interest to prove these properties starting, as far
as possible, from first principles. It is the purpose of this paper to provide
such a proof, in the framework of large-$N_c$ QCD. More precisely, we prove that
in the large-$N_c$ limit there exist stable hybrid baryons made up of heavy quarks.
Our proof is based on the assumption of color confinement and on the validity 
of a Hartree description for these states.

In Sec.~II we discuss the spin-flavor structure of the ground state hybrid 
baryons containing heavy quarks in large-$N_c$ QCD and show that it is closely 
related to that of orbitally excited baryons. 
Two types of hybrid baryons are considered: the first with symmetric orbital
wave functions and mixed symmetry spin-flavor content, and the second with a
mixed symmetric orbital wave functions and symmetric spin-flavor content.
The close connection between hybrid
baryons with symmetric orbital wave functions
and ordinary baryons is made explicit in
Sec.~III, where it is proven that the two types of baryons lead to the same Hartree
equations in the large-$N_c$ limit. Intuitively speaking, this is due to the
fact that the interquark forces in the two cases tend to the same value as
$N_c$ is taken to be large. The couplings of these states to mesons and
hybrid mesons are discussed in Sec.~IV and the corresponding $N_c$ counting rules
are established. These are used in Sec.~V to derive consistency conditions for
the spin-isospin structure of the couplings to mesons, which are then solved
explicitly. 

\section{Hybrid baryons with heavy quarks}

We will restrict ourselves in this paper to the tractable case of hybrid baryons
made up only of heavy quarks, for which the constituent quark picture becomes 
exactly valid. The wavefunction of the hybrid can be written as
\bea\label{1}
\Psi_{B_{\rm h}} = \frac{(-)^{\phi}}{\sqrt{N_c(N_c^2-1)}}\sum_{n=1}^{N_c^2-1}
\sum_{i=1}^{N_c} \chi^n(J_g,m_g)\otimes
[\psi^n_{a_1 a_2 \dots a_{N_c}}]_i\otimes \Psi_i(\vec x_1,\cdots,
\vec x_{N_c}; S,I;m_S,\alpha) CG\,,
\eea
where $\chi^n$ is the wavefunction of the gluon field which carries
angular momentum $J_g$. $\psi^n$ is the color part ($a_i$ are color indices 
for the quarks), $\Psi$ is the orbital part of the quarks' 
wavefunction and 
$CG$ is the Clebsch-Gordan coefficient $\langle Jm|SJ_g;m_S m_g\rangle $.

Since the hybrid is a color singlet, the quark part of the wavefunction
$\psi^n$ must transform in the adjoint representation of the color 
$SU(N_c)$ group. The normalized color part of the wavefunction reads
\bea\label{2}
[\psi^n_{a_1 a_2\dots a_{N_c}}]_i = \sqrt{\frac{2}{(N_c-1) !}}
\varepsilon_{a_1 a_2 \dots b_i \dots a_{N_c}}(t^n)_{a_i b_i}\,.
\eea
The $i^{th}$ quark will be called the external quark, and the remaining $N_c-1$
quarks will be called core quarks.
The Gell-Mann matrices $t^n$ are normalized
in the standard way by Tr$(t^n t^m)=1/2\delta_{nm}$.
These color wavefunctions are normalized as
\bea\label{psinorm}
\langle \psi^n_i | \psi^m_j \rangle &=& \delta_{nm}\quad (i=j)\,,\qquad
-\frac{1}{N_c-1} \delta_{nm}\quad (i\neq j)\,.
\eea
These relations and many others below can be obtained by making use of the 
identity
\beq
\sum_{x=1}^{N_c^2-1} t^x_{ij} t^x_{kl} = \frac12 \delta_{kj}\delta_{il} -
\frac{1}{2N_c}\delta_{ij}\delta_{kl}\,.
\eeq

The Pauli principle requires the hybrid wavefunction to be antisymmetric 
under any permutation of the quarks. This constrains the spin-flavor-orbital 
wavefunction to be described by a Young diagram which is the transpose of 
that corresponding to the color wavefunction (see Eq.~(\ref{fig1})). Therefore the
quarks in a hybrid can be effectively considered as identical particles
subjected to a special statistics: their wavefunction must have permutation 
symmetry described by the second Young diagram in Eq.~(\ref{fig1}).

\begin{eqnarray}\label{fig1}
\raisebox{-38.0pt}{\hbox{\rule{0.4pt}{10.0pt}\hskip-0.4pt%
\rule{10.0pt}{0.4pt}\hskip-10.0pt%
\rule[10.0pt]{10.0pt}{0.4pt}}\rule[10.0pt]{0.4pt}{0.4pt}\hskip-0.4pt%
\rule{0.4pt}{10.0pt}}\hskip-10.4pt
   \raisebox{-28pt}{\hbox{\rule{0.4pt}{10.0pt}\hskip-0.4pt%
\rule{10.0pt}{0.4pt}\hskip-10.0pt%
\rule[10.0pt]{10.0pt}{0.4pt}}\rule[10.0pt]{0.4pt}{0.4pt}\hskip-0.4pt%
\rule{0.4pt}{10.0pt}}\hskip-10.4pt
   \raisebox{-18pt}{\hbox{\rule{0.4pt}{10.0pt}\hskip-0.4pt%
\rule{10.0pt}{0.4pt}\hskip-10.0pt%
\rule[10.0pt]{10.0pt}{0.4pt}}\rule[10.0pt]{0.4pt}{0.4pt}\hskip-0.4pt%
\rule{0.4pt}{10.0pt}}\hskip-10.4pt
   \raisebox{-8pt}{\hbox{\rule{0.4pt}{10.0pt}\hskip-0.4pt%
\rule{10.0pt}{0.4pt}\hskip-10.0pt%
\rule[10.0pt]{10.0pt}{0.4pt}}\rule[10.0pt]{0.4pt}{0.4pt}\hskip-0.4pt%
\rule{0.4pt}{10.0pt}}\hskip-10.4pt
   \raisebox{2pt}{\hbox{\rule{0.4pt}{10.0pt}\hskip-0.4pt%
\rule{10.0pt}{0.4pt}\hskip-10.0pt%
\rule[10.0pt]{10.0pt}{0.4pt}}\rule[10.0pt]{0.4pt}{0.4pt}\hskip-0.4pt%
\rule{0.4pt}{10.0pt}}\hskip-0.4pt
   \raisebox{2pt}{\hbox{\rule{0.4pt}{10.0pt}\hskip-0.4pt%
\rule{10.0pt}{0.4pt}\hskip-10.0pt%
\rule[10.0pt]{10.0pt}{0.4pt}}\rule[10.0pt]{0.4pt}{0.4pt}\hskip-0.4pt%
\rule{0.4pt}{10.0pt}}
\quad\,,\quad
\overbrace{\,
\raisebox{-8.0pt}{\hbox{\rule{0.4pt}{10.0pt}\hskip-0.4pt%
\rule{10.0pt}{0.4pt}\hskip-10.0pt%
\rule[10.0pt]{10.0pt}{0.4pt}}\rule[10.0pt]{0.4pt}{0.4pt}\hskip-0.4pt%
\rule{0.4pt}{10.0pt}}\hskip-10.4pt
        \raisebox{2pt}{\hbox{\rule{0.4pt}{10.0pt}\hskip-0.4pt%
\rule{10.0pt}{0.4pt}\hskip-10.0pt%
\rule[10.0pt]{10.0pt}{0.4pt}}\rule[10.0pt]{0.4pt}{0.4pt}\hskip-0.4pt%
\rule{0.4pt}{10.0pt}}\hskip-0.4pt
\raisebox{2pt}{\hbox{\rule{0.4pt}{10.0pt}\hskip-0.4pt%
\rule{10.0pt}{0.4pt}\hskip-10.0pt%
\rule[10.0pt]{10.0pt}{0.4pt}}\rule[10.0pt]{0.4pt}{0.4pt}\hskip-0.4pt%
\rule{0.4pt}{10.0pt}}\,\cdots\,
\raisebox{2pt}{\hbox{\rule{0.4pt}{10.0pt}\hskip-0.4pt%
\rule{10.0pt}{0.4pt}\hskip-10.0pt%
\rule[10.0pt]{10.0pt}{0.4pt}}\rule[10.0pt]{0.4pt}{0.4pt}\hskip-0.4pt%
\rule{0.4pt}{10.0pt}}\,}^{N_c-1}
\end{eqnarray}

The quark color wavefunction (\ref{2}) is only one of many other possible
ways of constructing an adjoint from $N_c$ fundamental fields. The most
general construction involves first arranging $N_c-j$ quarks into an 
antisymmetric representation. The remaining $j$ quarks are then coupled 
either into an antisymmetric representation or a representation whose 
Young diagram has the same form as the first diagram in Eq.~(\ref{fig1}). 
Each of these possibilities could in principle generate one distinct set 
of degenerate states. This ambiguity in constructing a hybrid baryon state
is in contrast to the situation for ordinary baryons, where each subset 
of $j$ quarks is in a color antisymmetric state. However, all these
states different from (\ref{2}) disappear in the physical limit $N_c=3$. 
Therefore we will focus in the following only on the hybrid baryon 
states (\ref{2}).

It is plausible to assume that the ground state hybrids will have all the
quarks in a state with zero 
orbital angular momentum $L=0$; thus their orbital wavefunction is 
completely symmetric. This implies that the spin-flavor part of the
wavefunction must have mixed symmetry.
This is exactly the same as the spin-flavor wavefunction of an orbitally 
excited baryon with one quark in an excited state. Therefore the 
spin-flavor structure of the ground state hybrids can be read off 
immediately from \cite{PY1} (for two light flavors) and from \cite{PY2} 
(for three light flavors). 
Its wavefunction can be written in Hartree form as
\bea\label{Hartree}
\Psi_i = \Phi(\vec x_1)\Phi(\vec x_2)\cdots \Phi(\vec x_{N_c})\otimes
|S,I;m,\alpha\rangle_i\,,
\eea
with $\Phi$ one-particle wavefunctions and $|S,I;m,\alpha\rangle_i$ is the
spin-flavor state with mixed permutation symmetry constructed in \cite{PY1}.

For example, assuming that the constituent gluon field carries quantum
numbers $J^P=1^+$, the resulting states are classified into three
infinite towers of states with $|J-I|\leq \Delta$, $\Delta=0,1,2$.
Each tower is identified by a value of $\Delta$ and its states are
degenerate in the large-$N_c$ limit. One obtains in this way the sequence of 
states
\bea\label{Delta0}
(I,J^P) &=& (\frac12,\frac12^+)\,, (\frac32,\frac32^+)\,,
(\frac52,\frac52^+)\,,\cdots\qquad (\Delta=0)\\\label{Delta1}
& & (\frac12,\frac12^+)\,,(\frac12,\frac32^+)\,,(\frac32,\frac12^+)\,,\cdots
\qquad (\Delta=1)\\\label{Delta2}
& &(\frac12,\frac32^+)\,,(\frac12,\frac52^+)\,,(\frac32,\frac12^+)\,,\cdots
\qquad (\Delta=2)\,.
\eea
These states are linear combinations of the states (\ref{1}) with 
well-defined spin $S$. The corresponding relations are given below in
(\ref{Deltabasis}).

In addition to the states (\ref{Hartree}) we will consider also a different
type of hybrid baryons for which the quark wavefunction is given by
\bea\label{Hartree1}
\Psi'_i = 
\Phi(\vec x_1)\Phi(\vec x_2)\cdots \Psi(\vec x_i)
\cdots \Phi(\vec x_{N_c})\otimes
|I,m,\alpha\rangle\,.
\eea
The spin-flavor part $|I,m,\alpha\rangle$ is completely symmetric and the
orbital part has mixed symmetry. Here $\Psi(\vec x)$ is a one-body Hartree
wavefunction which will be assumed to be also $s$-wave. $\Psi(\vec x)$ can 
be taken to be orthogonal to $\Phi$ without any loss of generality.
Indeed, the component of $\Psi$ along the direction of $\Phi$ will give a
vanishing result when inserted into (\ref{Hartree1}) due to the identity
\bea
\sum_{i=1}^{N_c} \varepsilon_{a_1 a_2 \cdots b_i\cdots a_{N_c}} (t^n)_{a_i b_i}
= 0\,.
\eea
This can be proved by taking the square and summing over $a_i$ with the help of 
(\ref{psinorm}) which gives a vanishing result. These states are expected to
lie above the ones described by (\ref{Hartree}). Assuming the same quantum
numbers for the color octet gluon field, they form in the large $N_c$ limit
a $\Delta=1$ tower with the spin-flavor content (\ref{Delta1}).

\section{Existence}

The one-body Hartree wavefunction $\Phi(\vec x)$ in (\ref{Hartree}) can be
determined from the variational principle
\bea\label{variation}
\delta\frac{\langle \Psi_{B_{\rm h}}|{\cal H}|\Psi_{B_{\rm h}}\rangle}
{\langle \Psi_{B_{\rm h}}|\Psi_{B_{\rm h}}\rangle} = 0
\eea
where the Coulomb gauge Hamiltonian ${\cal H}$ is given by
\bea\label{Ham}
{\cal H} &=& \frac12\int\mbox{d} \vec x(\vec E_\perp^{a2} + \vec B^{a2}) +
\sum_{n}
\frac{1}{2m_Q}(-i\nabla_n-g\vec A^a(\vec x_n) t_n^a)^2 +
\frac{g^2}{4\pi}\sum_{m < n}\frac{t^x_m t^x_n}{|\vec x_m - \vec x_n|}\\
 &-&
\frac{g^2}{4\pi} f_{abc} \sum_n \int\mbox{d} \vec x
\frac{\vec A^b(\vec x) \cdot \vec E^c(\vec x) t_n^a}{|\vec x-\vec x_n|}
+ \frac{g^2}{8\pi} f_{abc}f_{ade}
\int\mbox{d} \vec x\mbox{d} \vec y
\frac{(\vec A^b\cdot \vec E^c)(\vec x) (\vec A^d\cdot \vec E^e)(\vec y)}
{|\vec x-\vec y|}\,.  \nonumber
\eea
The chromoelectric field is given by $\vec E^a=-\partial_0\vec A^a -
\nabla A^{0a}+gf_{abc}A^{0b}\vec A^c$ and includes, in addition to the
transverse part $\vec E^a_\perp$, also a longitudinal
component. In the expression for the Hamiltonian (\ref{Ham}) it is the full
chromoelectric field which appears, except in the first term.

We assume, as everywhere else, that the quarks are sufficiently heavy so that their
interactions are purely Coulombic. We neglected in (\ref{Ham}) terms
of higher order in the inverse heavy quark mass.
The expectation value of the color factor $t^x_m t^x_n$ in the state (\ref{1}), 
(\ref{Hartree}) can be easily computed with the result
\bea
{\cal C} = \frac{\langle \Psi_{B_{\rm h}}|\sum_x t^x_m t^x_n |\Psi_{B_{\rm h}}\rangle}
{\langle \Psi_{B_{\rm h}}|\Psi_{B_{\rm h}}\rangle} = \frac{-N_c^2+2N_c+1}{2N_c(N_c-1)}
\eea
for any $m\neq n$.
We made use here of the methods developed in \cite{PY1}
for computing matrix elements on the states with mixed symmetry $|SI\rangle_i$.
Keeping only the leading terms in $N_c$ we obtain in this way the following 
expression for the expectation value of the Hamiltonian (\ref{Ham})
\bea\label{var1}
\delta\left(\sum_{n=1}^{N_c}\int\mbox{d}\vec x_n\,
\Phi^\dagger(\vec x_n) H_0(\vec x_n) \Phi(\vec x_n) + \frac{g^2}{4\pi}{\cal C}\sum_{m < n}
\int\mbox{d}\vec x_n\mbox{d}\vec x_m\,
\frac{|\Phi(\vec x_n)|^2 |\Phi(\vec x_m)|^2}{|\vec x_m-\vec x_n|}\right) = 0
\eea
with
\bea\label{oneH}
 H_0 = -\frac{1}{2m_Q}\nabla^2 +
\frac{ig}{m_Q} \vec A(x)\cdot \nabla + \frac{g^2}{2m_Q}\vec A(x)\cdot 
\vec A(x) \,.
\eea
The terms proportional to $g$ and $g^2$ in (\ref{oneH}) describe the 
interaction of the  quarks with the transverse gluon field.
The $O(g)$ term appears to contribute a term of order $N_c g = 1/\sqrt{N_c}$ 
to the hybrid 
mass. However, its matrix elements taken on the $s$-wave Hartree wavefunctions
considered here vanish:
\bea
\int\mbox{d}\vec x \Phi(\vec x)\vec A\cdot \nabla\Phi(\vec x) =
\frac12 \int\mbox{d}\vec x \partial_i (\Phi(\vec x) A^i \Phi(\vec x)) -
\frac12 \int\mbox{d}\vec x \Phi(\vec x) (\nabla\cdot\vec A) \Phi(\vec x) = 0\,.
\eea
Therefore the leading correction to the hybrid baryon mass comes from the ``seagull''
term in (\ref{oneH}) (quadratic in $A^a$) and is of order $g^2 N_c = N_c^0$.
Also, the instantaneous
Coulomb quark-gluon interactions in (\ref{Ham}) contribute only
to order $g^2 N_c = N_c^0$ to (\ref{var1}) and is unimportant for large $N_c$.

One can see thus that the
large $N_c$ limit of the hybrid baryon Hartree equation (\ref{var1})
contains only the kinetic quark terms and the quark-quark Coulomb interactions,
just as for ordinary baryons. Furthermore, for large $N_c$ this equation
is exactly identical to the corresponding equation
for an ordinary baryon \cite{Wi} (for which the color factor 
${\cal C}$ takes the value $-(N_c+1)/(2N_c)$).

We conclude therefore that the one-particle Hartree wavefunctions 
$\Phi(\vec x)$ in (\ref{Hartree}) are the same as in 
ordinary baryons. Hence the mass of a hybrid baryon grows linearly
with $N_c$ in the same 
way as for an ordinary baryon, as expected on intuitive grounds. 
The different expectation values of the gluon field energy and gluon-quark 
interactions in (\ref{Ham}) produce a finite mass difference among these types 
of states of order $N_c^0$. This mass difference can be interpreted 
phenomenologically as a  constituent gluon mass.

Turning to the applicability of these results to the physical world
with $N_c=3$, one can see that the Coulomb quark-quark
interaction in a hybrid is attractive for any $N_c\geq 3$. This
suggests the existence of these states also in the physical case $N_c=3$.
However, for finite values of $N_c$ the quark-gluon interactions must
be taken into account too.
Calculations in the bag model \cite{1,2} show that these additional 
interaction terms are likely to be small and can be treated as a 
perturbation.

The Hartree equation for the hybrid with orbital wavefunction with mixed
symmetry (\ref{Hartree1}) has the form
\bea\label{var2}
& &\qquad\qquad
\delta\left\{(N_c-1)\int\mbox{d}\vec x\,\Phi^\dagger(\vec x) H_0 \Phi(\vec x)
+ \int\mbox{d}\vec x\,\Psi^\dagger(\vec x) H_0 \Psi(\vec x)\right.\\
& &\left.+g^2\frac{N_c(N_c-1)}{2}\left(
-\frac{(N_c+1)(N_c-2)}{2N_c^2}
\int\mbox{d}\vec x_m\mbox{d}\vec x_n\frac{|\Phi(\vec x_m)|^2 |\Phi(\vec x_n)|^2}
{4\pi|\vec x_m-\vec x_n|}\right.\right.\nonumber\\
& &\qquad\qquad\left.\left. +
\frac{1}{N_c^2(N_c-1)}
\int\mbox{d}\vec x_m\mbox{d}\vec x_n\frac{|\Phi(\vec x_m)|^2 |\Psi(\vec x_n)|^2}
{4\pi|\vec x_m-\vec x_n|}\right.\right.\nonumber\\
& &\qquad\qquad\left.\left. -
\frac{N_c^2-N_c-1}{N_c^2(N_c-1)}
\int\mbox{d}\vec x_m\mbox{d}\vec x_n\frac{(\Phi^\dagger\Psi)(\vec x_m) 
(\Psi^\dagger\Phi)(\vec x_n)}
{4\pi|\vec x_m-\vec x_n|}\right)\right.\nonumber\\
& &\left.-\frac{1}{2(N_c^2-1)}\langle\chi^y|
(if_{yax}-d_{yax}) 
\int\mbox{d}\vec x\mbox{d}\vec x_n\frac{g^2f_{abc}(\vec A^b\cdot\vec E^c)(\vec x)
|\Phi(\vec x_n)|^2 }{4\pi|\vec x-\vec x_n|}|\chi^x \rangle\right.\nonumber\\
& &\left.-\frac{1}{2(N_c^2-1)}\langle\chi^y|
(if_{yax}+d_{yax}) 
\int\mbox{d}\vec x\mbox{d}\vec x_n\frac{g^2f_{abc}(\vec A^b\cdot\vec E^c)(\vec x)
|\Psi(\vec x_n)|^2 }{4\pi|\vec x-\vec x_n|}|\chi^x\rangle
\right\} = 0\nonumber\,.
\eea
We included here the quark-quark and quark-gluon Coulomb interaction terms. From 
this equation it is easy to read off the couplings of the quarks with each
other and with the gluon field. 
The $N_c-1$ quarks in the core interact with each other by Coulomb interaction
to leading order and with the gluon field at subleading order in $1/N_c$. Hence
the latter term can be neglected in the equation for $\Phi$, which becomes
thus identical to (\ref{var1}). Its solution is identical to the Hartree
wavefunction in an ordinary baryon.

The dynamics of the external quark is more complex: it feels a repulsive
force from the $N_c-1$ quarks in the core of the order $1/N_c^2$ (the term in  the
third line of (\ref{var2})) and an exchange interaction of order 1
(the fourth line of (\ref{var2})) which can be either attractive or repulsive. 
However, the hybrid state can still be bound because
the external quark interacts also with the gluon field to order 1 (the last
term in (\ref{var2})). We can estimate the nature of this interaction by
assuming that the gluon field is classical (this is esentially the assumption
made in bag model calculations \cite{1,2}). Then the matrix elements of the
gluon fields can be written in terms of a mode function $\chi(\vec x)$ as
\bea\label{Amatel}
& &\langle 0|\vec A^a(\vec x)|\chi^n(J_g m_g)\rangle =
\vec\varepsilon\,(m_g,\hat x) \chi(|\vec x |) \mbox{e}^{-i\omega t} \delta_{na}\\
\label{Ematel}
& &\langle 0|\vec E^a(\vec x)|\chi^n(J_g m_g)\rangle =
i\omega \vec\varepsilon\,(m_g,\hat x) \chi(|\vec x |) \mbox{e}^{-i\omega t} \delta_{na}\,.
\eea
The matrix element of the field products in (\ref{var2}) can be written in terms
of these functions with the help of a variant of ``vacuum factorization''
\bea
\langle\chi^y(m_g)|\vec A^a(\vec x)\cdot\vec E^b(\vec x)|\chi^x(m_g)\rangle
= i\omega |\chi(\vec x)|^2 (\delta_{yb}\delta_{cx} - \delta_{yc}\delta_{bx})\,.
\eea
Introducing this expression into (\ref{var2}) one obtains the following expectation
value of the Coulomb quark-gluon interaction
\bea\label{qu-gl}
\langle\Psi'_{B_{\rm h}}|H_{q-g}|\Psi'_{B_{\rm h}}\rangle =
- N_c g^2\omega\left(
\int\mbox{d}\vec x\mbox{d}\vec x_n\frac{|\chi(\vec x)|^2
|\Phi(\vec x_n)|^2 }{4\pi|\vec x-\vec x_n|} +
\int\mbox{d}\vec x\mbox{d}\vec x_n\frac{|\chi(\vec x)|^2
|\Psi(\vec x_n)|^2 }{4\pi|\vec x-\vec x_n|}\right)\,,
\eea
which is seen to be attractive\footnote{The quark-gluon Coulomb interaction in the
state (\ref{Hartree}) has exactly the same form as the first term in (\ref{qu-gl}).}.

The properties of these states following from their Hartree description
are very similar to those of the hybrids with symmetric orbital wavefunction
(\ref{Hartree}). Their mass grows linearly with $N_c$ in the same way as
for ordinary baryons. The interaction energy with the gluon field and
the energy of the external quark are responsible for a mass splitting among
these states of order $N_c^0$.

It is interesting to compare the hybrid baryons to the hybrid mesons,
recently discussed in the large $N_c$ limit in \cite{Co}.
Let us consider, for illustration, a hybrid meson made up of a heavy 
quark-antiquark pair. Its wavefunction can be written analogously to (\ref{1})
\bea\label{hmeson}
\Psi_{M_{\rm h}} = \frac{1}{\sqrt{N_c^2-1}}\sum_{n=1}^{N_c^2-1}\chi^n\otimes
\psi^n_{a_1 a_2}(\vec x_1,\vec x_{2})\,.
\eea
The color part of the quarks' wavefunction is 
$\psi^n_{a_1 a_2}=\sqrt2 (t^n)_{a_1 a_2}$,
which gives the following color-factor for the quark-antiquark potential 
in such a bound state
\bea\label{10}
\langle \psi^n| t_1^x \bar t_2^x | \psi^m\rangle = \frac{1}{2N_c}\delta_{nm}
\qquad (\bar t^x=-(t^x)^T)\,.
\eea
This produces a repulsive quark-antiquark force which vanishes in the large
$N_c$ limit. For comparison, the value of the corresponding color
factor in a color-singlet meson state 
$\psi=\frac{1}{\sqrt{N_c}}\delta_{a_1 a_2}$ is
\bea
\langle \psi| t_1^x \bar t_2^x | \psi\rangle = -\frac{N_c^2-1}{2N_c}
\eea
which gives the usual attractive potential between a quark and antiquark 
in a meson. 

These results show that if hybrid mesons are to exist in the large $N_c$
limit (as shown in \cite{Co}) then the color-octet gluon field must
supply the attractive force needed to bind the system. Furthermore, in the
physical case $N_c=3$ this quark-gluon attractive interaction must be strong
enough to overcome the finite Coulomb quark-antiquark repulsion.
This is in contrast to the case of the hybrid baryons (\ref{Hartree}), where the 
dynamical effect of the  color-octet gluon field on the quarks
can be neglected for large $N_c$ and the
quark-quark interaction is attractive for any $N_c$ larger than 2.
The hybrid mesons are somewhat similar to the hybrid baryons (\ref{Hartree1}),
in which the external quark is repelled by the core with a strength $1/N_c^2$.
However, there is no analog for the exchange interaction term in hybrid baryons
for the hybrid meson case.

We have not included here the contribution to the total energy (\ref{var1})
and (\ref{var2}) from the gluon field. In the same approximation as in
(\ref{Amatel}) and (\ref{Ematel}), one simply replaces the field operators
in (\ref{Ham}) by the corresponding classical functions. The attractive
quark-gluon interactions should localize the gluon field within a region
of size $1/\Lambda_{QCD}$ surrounding the quarks. Suffice it to note that
these contributions are all of order $N_c^0 \Lambda_{QCD}$.

To complete our proof of existence of the hybrid baryons, we still have to
show that these states are relatively narrow. It is known that meson coupling
to baryons can grow with $N_c$. This mechanism could potentially cause the 
widths of these states to grow with $N_c$ too, which would effectively render 
them unstable. We will show in the following section by means of computations
in the quark model that some of these states are indeed narrow in the 
large-$N_c$ limit. 

\section{Quark model predictions for hybrid baryons}

The Hartree picture of a hybrid baryon developed in Sect.~II can be used in the 
same way as for ordinary baryons \cite{Wi} to derive scaling laws for its 
couplings to mesons. The properties of the two types of hybrid baryons
(\ref{Hartree}) and (\ref{Hartree1}) will be seen to be different.
We will consider them in the following in turn.

\subsection{Hybrid baryons with symmetric orbital wavefunction}

The coupling of a pion to a hybrid baryon (\ref{Hartree}) is parametrized in 
terms of the matrix element
\bea\label{Z}
& &\langle S'I'J_g;m'_S\alpha' m'_g
|\sum_{n=1}^{N_c} \sigma_n^i\tau_n^a |
SIJ_g;m_S\alpha m_g\rangle =\\
& &\frac{1}{N_c(N_c^2-1)}(-)^{\psi(S'I'i')+\psi(SIi)}
\delta_{m_g m'_g}
\sum_{y=1}^{N_c^2-1}
\sum_{k,k'=1}^{N_c} \,_{k'}\langle S'I';m'_S\alpha'|\sum_{n=1}^{N_c} \sigma_n^i\tau_n^a|
SI;m_S\alpha\rangle_k\, (\bar\psi^y_{k'}\psi^y_k) =\nonumber\\
& &\qquad (-)^{\psi(S'I'i')+\psi(SIi)}\delta_{m_g m'_g}\nonumber\\
& &\qquad\times 
\left\{ \,_{k}\langle S'I';m'_S\alpha'|\sum_{n=1}^{N_c} \sigma_n^i\tau_n^a|
SI;m_S\alpha\rangle_k\,
- \,_{k'}\langle S'I';m'_S\alpha'|\sum_{n=1}^{N_c} \sigma_n^i\tau_n^a|
SI;m_S\alpha\rangle_k\right\}\,.\quad (k\neq k')\nonumber
\eea
The color part of the matrix element has been computed with the help of the
relations (\ref{psinorm}). The matrix elements appearing in the last line
have been computed already (see Eqs.~(4.81) and (4.82) in \cite{PY1}). 
The ``physical'' value of the matrix element is obtained, just as in 
ordinary baryons, after dividing with the square roots of the norms of 
the initial and final states
\bea
\langle SIJ_g|SIJ_g\rangle = 3(2i+1)
\left\{ \begin{array}{ccc}
S & I & 1 \\
\frac12 & \frac12 & i \end{array}\right\}^2\,.
\eea
We obtain finally the following result for the matrix element (\ref{Z})
\bea\label{Zfin}
& &\langle S'I'J_g;m'_S\alpha' m'_g|\sum_{n=1}^{N_c} \sigma_n^i\tau_n^a |
SIJ_g;m_S\alpha m_g\rangle =\\
& &\quad N_c (-)^{S+I'}\sqrt{(2S+1)(2I+1)}
\left\{ \begin{array}{ccc}
S' & S & 1 \\
I & I' & 1 \end{array}\right\}
\langle S'm'_S|S1;m_S i\rangle \langle I'\alpha'|I1;\alpha a\rangle 
\delta_{m_g m'_g}\,.\nonumber
\eea
We used here the same phase as in \cite{PY1} for the mixed symmetry states
$\psi(SIi)=i+I+\frac12$.

After dividing with the meson decay constant $f_M\propto N_c^{\frac12}$ this gives 
that mesons couple to hybrid baryons with a coupling of order $N_c^{\frac12}$,
just as to ordinary baryons. It is interesting to note that the quark model result 
(\ref{Zfin}) coincides exactly with the one obtained in \cite{PY1} for pion 
couplings to orbitally excited baryons.

It is easy to see that the decay mode $B_{\rm h} \to M + B$ is forbidden for
hybrid baryons made up of heavy quarks. In this approximation the transition
matrix element is proportional to the overlap of the color-singlet and 
color-octet wavefunctions respectively
$\langle \psi|\psi^n_i\rangle = 0$. On the other hand, the lowest lying
hybrid baryons can decay to a pair $(M_{\rm h},B)$ in a $p$-wave,
with $M_{\rm h}$ a hybrid meson and $B$ an ordinary $s$-wave baryon.

The decays to an ordinary meson and baryon are induced through mixing
$M_{\rm h} \leftrightarrow M$. This mixing is present already at order 1 in the $1/N_c$
expansion, unless the respective state $M_{\rm h}$ has exotic quantum numbers.
Therefore the scaling laws for the hybrid meson couplings are the same as
those for ordinary mesons.
As mentioned above, the mixing $M_{\rm h} \leftrightarrow M$ vanishes in the heavy 
quark limit as $N_c^0/m_Q$. This has been taken into account in recent refinements
of quarkonium physics, where a small octet $\bar QQg$ admixture has been included 
in addition to the $\bar QQ$ component \cite{BBL}.

In the following we will compute the decay amplitudes of a hybrid
baryon in the quark model. The detailed form of the respective coupling
depends on the quantum numbers of the final hybrid meson and is shown
below for a few cases of physical interest.
\bea\label{couplings}
{\cal L}_{\rm int} &=& \frac{g_1}{f_M}
(\bar q\gamma_\mu \tilde F_{\mu\nu}\tau^a q)\,
\partial_\nu M^a(0^{-+})
+ \frac{g_2}{f_M}(\bar q\gamma_\mu\gamma_5 F_{\mu\nu}\tau^a q)\,
\partial_\nu M^a(0^{--})\\
&+& \frac{g_3}{f_M}(\bar q F_{\mu\nu}\tau^a q) \,\partial_\mu M_\nu^a(1^{--})
+ \frac{g_4}{f_M}
(\bar q \sigma_{\mu\nu} F_{\nu\lambda}\tau^a q)\,\partial_\lambda M_\mu^a(1^{-+})
\nonumber\\
&+& \frac{g_5}{2f_M}
[\bar q (\gamma_\mu F_{\alpha\nu} + \gamma_\nu F_{\alpha\mu})\gamma_5\tau^a q]\,
\partial_\alpha M_{\mu\nu}^a(2^{--})\,.\nonumber
\eea
The quantum numbers $J^{PC}$ of the respective hybrid meson state are shown in the 
brackets.
We will be interested in the following only in decays $B_{\rm h}\to B+M_{\rm h}$ where both
$B_{\rm h}$ and $B$ have positive parity. Then an odd parity hybrid meson $M_{\rm h}$ is emitted 
in an orbital $p$-wave. The decay amplitudes for these transitions can be extracted
from (\ref{couplings}) and are given, in the nonrelativistic limit, by
\bea\label{Ma}
{\cal M}_{\rm a} &=& \langle B|\sum_{n=1}^{N_c}(\vec p\cdot
g\vec B^x(\vec r_n) ) t^x_n \tau^a_n| B_{\rm h}\rangle\qquad\qquad\qquad\qquad\quad
(J^{PC}=0^{-+})\\\label{Mb}
{\cal M}_{\rm b} &=& \langle B|\vec p\cdot \sum_{n=1}^{N_c}(\vec\sigma
\times g\vec B^x(\vec r_n) ) t^x_n \tau^a_n| B_{\rm h}\rangle\qquad\qquad\quad\qquad
(J^{PC}=0^{--})\\\label{Mc}
{\cal M}_{\rm c} &=& \langle B|(\vec p\times\vec\varepsilon) \cdot
\sum_{n=1}^{N_c}
g\vec B^x(\vec r_n)  t^x_n \tau^a_n| B_{\rm h}\rangle\qquad\qquad\qquad\quad (J^{PC}=1^{--})\\
\label{Md}
{\cal M}_{\rm d} &=& (\delta_{ij}(\vec p\cdot\vec\varepsilon) 
-\varepsilon_i p_j) \langle B|
\sum_{n=1}^{N_c}
gB^{i,x}(\vec r_n)\sigma^j_n  t^x_n \tau^a_n| B_{\rm h}\rangle\qquad (J^{PC}=1^{-+})\\
\label{Me}
{\cal M}_{\rm e} &=& 2\epsilon_{mqi}\varepsilon_{jq} p_m\,
\langle B| \sum_{n=1}^{N_c}
gB^{i,x}(\vec r_n)\sigma^j_n  t^x_n \tau^a_n| B_{\rm h}\rangle\qquad\qquad\,\,\,
 (J^{PC}=2^{--})\,.
\eea

The matrix elements ${\cal M}_{\rm a}$ and  ${\cal M}_{\rm c}$ are related in a
simple way. They can be computed with the methods of \cite{PY1} with the result
\bea\label{Mares}
& &\langle I',m'\alpha'| \sum_{n=1}^{N_c}gB^{i}(\vec r_n) \tau_n^a|
(SJ_g)JI,m\alpha\rangle = \\
& &\quad \frac{1}{\sqrt6}N_c (-)^{S-J+1+2I}\frac{\sqrt{(2I+1)(2J+1)}}{2I'+1}
\langle 0|\!|gB|\!|\chi\rangle\, 
\delta_{SI'}\,
\langle I'm'|J1;mi\rangle \langle I'\alpha'|I1;\alpha a\rangle \nonumber\,.
\eea
The reduced matrix element of the chromomagnetic field $\vec B^a$ is defined 
by
\bea
\int \mbox{d}\vec r\, \Phi^\dagger(\vec r\,)\Phi(\vec r\,)
\langle 0| gB^a_i(\vec r\,)|\chi^n(J_g,m_g)\rangle = 
\langle 0|\!| gB|\!|\chi\rangle \langle 0|J_g,1; m_g,i\rangle \delta_{an}
\eea

The remaining matrix elements (\ref{Mb}), (\ref{Md}) and (\ref{Me})
can be expressed in terms of the following basic quantity.
\bea\label{MT}
& &{\cal M}^{T}_\ell = \langle T\ell |11;ji\rangle 
\langle I',m'\alpha'| 
\sum_{n=1}^{N_c}gB^{i,x}(\vec r_n)\sigma^j_n  t^x_n \tau^a_n|(SJ_g)JI,m\alpha\rangle
=\\
& &\quad N_c (-)^{2J-I+S+T+1}\sqrt{\frac{(2I+1)(2J+1)(2S+1)(2T+1)}{2I'+1}}
\left\{ \begin{array}{ccc}
1 & 1 & T \\
J & I' & S \end{array}\right\}
\left\{ \begin{array}{ccc}
1 & 1 & 1 \\
S & I & I' \end{array}\right\}\nonumber\\
& &\qquad \times\langle 0|\!| gB|\!|\chi\rangle\,
\langle I'm'|JT;m\ell\rangle \langle I'\alpha'|I1;\alpha a\rangle \nonumber\,.
\eea

The amplitude ${\cal M}_{\rm b}$ is given by
\bea
{\cal M}_{\rm b} = -i\sqrt2\sum_\ell (-)^\ell {\cal M}^{T=1}_\ell  p^{-\ell}\,.
\eea

The matrix element (\ref{Md}) can be written most
conveniently in terms of a tensor $t(T,\ell)$ defined by
\bea\label{tdef}
t(T,\ell) = \langle T\ell| 11;ij\rangle p^i\varepsilon^j\,,\qquad\qquad
p^i\varepsilon^j = \sum_{T,\ell}t(T,\ell) \langle T\ell| 11;ij\rangle\,.
\eea
Then one has
\bea
{\cal M}_{\rm d} = 2{\cal M}^{T=0}t(0) + \sum_\ell (-)^\ell
{\cal M}^{T=1}_\ell t(1,-\ell)
- \sum_\ell (-)^\ell {\cal M}^{T=2}_\ell t(2,-\ell)\,.
\eea

The advantage of writing the decay amplitude in this form is that the angular
distribution of the emitted vector mesons can be expressed in a simple form.
For example, for a general amplitude
\bea
{\cal M}(\vec p,\vec \varepsilon) = \sum_{T=0,1,2}c_T \sum_\ell (-)^\ell
t(T,-\ell) \langle J'm'|JT;m\ell\rangle
\eea
one obtains the following spin-averaged angular distribution
\bea
\Gamma \simeq \frac{1}{2J+1}\sum_{m,m'}|{\cal M}(\vec p,\vec \varepsilon) |^2 =
\frac{2J'+1}{2J+1}\sum_T \frac{|c_T|^2}{2T+1} \sum_\ell |t(T,\ell)|^2
\eea
with 
\bea
\sum_\ell |t(T,\ell)|^2 &=& \frac13|\vec p\cdot\vec\varepsilon\,|^2 \qquad\qquad 
\qquad\,\,\,(T=0)\\
&=& \frac12|\vec p\times \vec\varepsilon\,|^2 \qquad\qquad\qquad (T=1)\\
&=& \frac23|\vec p\cdot\vec\varepsilon\,|^2 + \frac12|\vec p\times \vec\varepsilon\,|^2
\quad\,\, (T=2)\,.
\eea
It should be noted that the 3-dimensional scalar product $\vec p\cdot\vec\varepsilon$
does not vanish for a longitudinally polarized vector meson.

Finally, the matrix element appearing in the amplitude (\ref{Me}) can be expressed
in terms of (\ref{MT}) as
\bea
\langle B| \sum_{n=1}^{N_c}
gB^{i,x}(\vec r_n)\sigma^j_n  t^x_n \tau^a_n| B_{\rm h}\rangle =
\sum_{T,\ell} {\cal M}^{T}_\ell \langle T\ell|11;ij\rangle\,.
\eea

The amplitudes for decays into isoscalar hybrid mesons can be computed
analogously with the help of the basic matrix elements
\bea
& &\langle I',m'\alpha'| \sum_{n=1}^{N_c}gB^{i}(\vec r_n)|
(SJ_g)JI,m\alpha\rangle = 0\\\label{M'}
& &{\cal M}^{'T}_\ell = \langle T\ell |11;ji\rangle 
\langle I',m'\alpha'| 
\sum_{n=1}^{N_c}gB^{i,x}(\vec r_n)\sigma^j_n  t^x_n|(SJ_g)JI,m\alpha\rangle
=\\
& &\quad N_c \frac{1}{\sqrt6}
(-)^{T+1}\sqrt{\frac{(2J+1)(2S+1)(2T+1)}{2I'+1}}
\left\{ \begin{array}{ccc}
1 & 1 & T \\
J & I' & S \end{array}\right\}
 \langle 0|\!| gB|\!|\chi\rangle\, \nonumber\\
& &\qquad\qquad\times 
\langle I'm'|JT;m\ell\rangle \delta_{II'}\delta_{\alpha\alpha'} \nonumber\,.
\eea

After dividing by the decay constant $f_{M_{\rm h}(M)}\propto N_c^{\frac12}$, the
quark model calculation shows therefore that the 
width of these states due to the decay mode $B_{\rm h}\to B+M_{\rm h}(M)$
is of order  $N_c^{0}$ in the large-$N_c$ limit.
This would be true if the mass difference
$M_{B_{\rm h}}-M_B$ was a constant in the same limit. 
Unfortunately, this turns out not to be the case due to a large
mixing between the ordinary baryons and the hybrid baryons. This mixing appears
first at order $1/m_Q$ and is induced by the chromomagnetic
term in the nonrelativistic expansion. The matrix element of this operator
is given by
\bea\label{mix}
& &\langle I'm'\alpha'|\frac{g}{2m_Q}\sum_{n=1}^{N_c}\vec\sigma_n\cdot\vec B(\vec r_n)|
(SJ_g)JI,m\alpha\rangle =\\
& &\qquad \frac{1}{2m_Q} \frac{N_c}{\sqrt6} (-)^{1+J+S}\sqrt{\frac{2S+1}{2J+1}}\delta_{JI}
\delta_{JI'}\, \langle 0|\!| gB|\!|\chi\rangle\,
\delta_{mm'}\delta_{\alpha\alpha'}\,.\nonumber
\eea
We made use here of the result (\ref{M'}) (with $T=0$) for the matrix element of an 
isoscalar operator. The mixing amplitude grows with $N_c$ 
like $\sqrt{N_c}/m_Q$ which is to be contrasted  
with the corresponding meson-hybrid meson mixing which is of order $N_c^0/m_Q$. 
Assuming the existence of only two such baryon states, it would appear that
the physical states are given in the large $N_c$ limit by the ``ideal-mixing''
combinations
\bea
|B_{1,2}\rangle = \frac{1}{\sqrt2}(|B\rangle \pm |B_{\rm h}\rangle )\,.
\eea
An infinite mixing would also repel the states by an infinite 
amount, such that the mass separation $M_{B_{\rm h}}-M_B$ grows like $N_c^{\frac12}$.
A $p$-wave decay width will be of order $N_c^{\frac32}$. We conclude thus that
some of the ground state hybrids (\ref{Hartree}) are unstable in the 
large-$N_c$ limit against $1/m_Q$ corrections 
(see the discussion following Eq.~(\ref{75}) in Sect.~V). 
Whether this type of states exists for $N_c=3$ is
an open question which we cannot answer.

\subsection{Hybrid baryons with mixed symmetry orbital wavefunction}

The mixing of the states (\ref{Hartree1}) with the ordinary baryons $Q^3$ 
induced at order 
$1/m_Q$ is parametrized by a matrix element analogous to (\ref{mix})
\bea\label{mix'}
\langle I'm\alpha|\frac{g}{2m_Q}\sum_{n=1}^{N_c}\vec\sigma_n\cdot\vec B(\vec r_n)|
JI;m\alpha\rangle
= \frac{1}{2m_Q} (-)\sqrt{\frac23} \sqrt{I(I+1)}\delta_{JI} \delta_{II'}
\, \langle 0|\!| gB|\!|\chi\rangle\,.
\eea
One can see that the mixing amplitude of these states with the ordinary
baryons vanishes in the large $N_c$ limit. Thus these states can be expected
to be stable with respect to mixing in this limit, in contrast
to the ground state hybrids (\ref{Hartree}).

The calculation of the couplings of the hybrid states (\ref{Hartree1})
proceeds in an analogous way. Mesons couple to them with a strength
of order $N_c^{\frac12}$, just as to the states (\ref{Hartree}).
The corresponding matrix element is
\bea\label{Z'}
& &\langle I'J_g;m'_S\alpha' m'_g|\sum_{n=1}^{N_c} \sigma_n^i\tau_n^a |
IJ_g;m_S\alpha m_g\rangle =\\
& &\qquad
\frac{1}{N_c(N_c^2-1)}\delta_{m_g m'_g}
\sum_{y=1}^{N_c^2-1}
\sum_{k,k'=1}^{N_c} \langle I'm'_S\alpha'|\sum_{n=1}^{N_c} \sigma_n^i\tau_n^a|
Im_S m_g\rangle\, (\bar\psi^y_{k'}\psi^y_k)\, \delta_{kk'} =\nonumber\\
& &\qquad 
\langle I'm'_S\alpha'|\sum_{n=1}^{N_c} \sigma_n^i\tau_n^a|Im_S\alpha\rangle
= N_c \sqrt{\frac{2I+1}{2I'+1}}
\langle I'm'_S|I1;m_S i\rangle \langle I'\alpha'|I1;\alpha a\rangle 
\delta_{m_g m'_g}\,.\nonumber
\eea

The decay amplitude of these hybrids into ordinary baryons plus (isovector) 
mesons are given by the following two basic matrix elements
\bea\label{Mares'}
& &\langle I',m'\alpha'| \sum_{n=1}^{N_c}gB^{i}(\vec r_n) \tau_n^a|
JI,m\alpha\rangle = \\
& &\quad \sqrt{\frac23} (-)^{I-J+1}\sqrt{I(I+1)}
\sqrt{\frac{2J+1}{2I+1}}
\langle 0|\!|gB|\!|\chi\rangle\,
\delta_{II'}\,
\langle I'm'|J1;mi\rangle \langle I'\alpha'|I1;\alpha a\rangle \nonumber
\eea
and
\bea\label{MTres'}
& &{\cal M}^{T}_\ell = \langle T\ell |11;ji\rangle 
\langle I',m'\alpha'| 
\sum_{n=1}^{N_c}gB^{i,x}(\vec r_n)\sigma^j_n  t^x_n \tau^a_n|JI,m\alpha\rangle
=\\
& &\quad \frac{1}{\sqrt6}N_c (-)^{2J-I'+I+T+1}
\sqrt{\frac{(2I+1)(2J+1)(2T+1)}{2I'+1}}
\left\{ \begin{array}{ccc}
1 & 1 & T \\
J & I' & I \end{array}\right\}\nonumber\\
& &\qquad \times\langle 0|\!| gB|\!|\chi\rangle\,
\langle I'm'|JT;m\ell\rangle \langle I'\alpha'|I1;\alpha a\rangle \nonumber\,.
\eea
The reduced matrix element of the chromomagnetic field $\vec B^a$ is defined 
for this case by
\bea
\int \mbox{d}\vec r\, \Psi^\dagger(\vec r\,)\Phi(\vec r\,)
\langle 0| gB^a_i(\vec r\,)|\chi^n(J_g,m_g)\rangle = 
\langle 0|\!| gB|\!|\chi\rangle \langle 0|J_g,1; m_g,i\rangle \delta_{an}
\eea

These results show that the decay widths of a hybrid baryon 
$\Psi'_{B_{\rm h}}$ into an ordinary baryon and an isovector meson with 
quantum numbers $J^{PC}=0^{-+}\,,1^{--}$
are of order $1/N_c^2$ (see (\ref{Mares'}), (\ref{Ma}) and (\ref{Mc})). 
On the other hand, the decays into mesons with quantum
numbers $J^{PC}=0^{--}\,,1^{-+}\,,2^{--}$ are enhanced and have widths of 
order $N_c^0$ (see (\ref{MTres'})). These couplings have the same $N_c$ 
scaling as the couplings of
the orbitally excited baryons with mixed symmetry discussed in \cite{PY1}.
In the following section we will abstract these results of the quark model
by keeping only their large $N_c$ scaling law to derive constraints on the
spin-flavor structure of the couplings.

\section{Consistency conditions for hybrid baryon couplings}

In this section we derive a set of model-independent constraints on the
hybrid baryons' coupling to mesons using the method of consistency conditions
(see \cite{DJM} and references therein).
These constraints arise from a mismatch in the large $N_c$ scaling laws for
scattering amplitudes, computed in quark diagram and hadron diagram language
respectively. For the latter one uses the scaling laws for vertices derived in
the quark model, with an implicit assumption that they still hold if the
quarks become light. We will study in the following the consistency conditions
obtained from the scattering amplitude $B_{\rm h}+M\to B+M_{\rm h}$.

There are two different classes of diagrams contributing to this scattering 
amplitude to first order in $\alpha_s$. Typical diagrams are shown in Fig.~1. 
Using Eq.~(\ref{1}) and the wavefunction of a hybrid meson (\ref{hmeson}), 
the dependence on  $N_c$ of these two contributions can be made explicit 
\bea\label{Ia}
I_1 &=& \frac{\alpha_s}{N_c(N_c^2-1)}\sum_{x,y=1}^{N_c^2-1}
\sum_{i,j=1}^{N_c} \mbox{Tr }[t^a \sqrt2 t^x]\, [\bar\psi t^a_j\psi^y_i]
\langle\chi^x(M_{\rm h})|\chi^y(B_{\rm h})\rangle {\cal M}^1_{i}(p)\\\label{Ib}
I_2 &=& \frac{\alpha_s}{N_c(N_c^2-1)}\sum_{x,y=1}^{N_c^2-1}\left\{
\sum_{i=k,j=1}^{N_c} [\bar\psi t^a_j t^a_k\sqrt2 t^x_k \psi^y_i]
\langle\chi^x(M_{\rm h})|\chi^y(B_{\rm h})\rangle {\cal M}^2_{i}(p)\right.\\
& &+\left. \sum_{i\neq k,j=1}^{N_c} [\bar\psi t^a_j t^a_k\sqrt2 t^x_k\psi^y_i]
\langle\chi^x(M_{\rm h})|\chi^y(B_{\rm h})\rangle {\cal M}^3_{i}(p)\right\}\nonumber
\quad(j\neq k)
\eea
where ${\cal M}^{1,2,3}_{i}(p)$ depend only on the momenta of the involved particles
but not on $N_c$. 
In (\ref{Ib}) one must sum only over terms with $j\neq k$ because only these
digrams can transfer momentum from the hybrid baryon to the meson. Furthermore, 
we took into account the fact that the momentum-dependent part of the diagram 
is different when the exchanged quark is identical or not with the external 
quark in the
hybrid. For the hybrids (\ref{Hartree}) this difference appears because of the
spin-flavor structure of the $|SI\rangle_i$ state and for the states 
(\ref{Hartree1}) it is due to the different orbital wavefunction of the 
external $i^{th}$ quark.

The overlap of the gluonic wavefunctions is given by
$\langle\chi^x(M_{\rm h})|\chi^y(B_{\rm h})\rangle={\cal I}(p)\delta_{xy}$ with ${\cal I}(p)$
another function of the hybrid momenta. The expressions (\ref{Ia}), (\ref{Ib}) can
be easily computed with the help of the relations (here $i$ is kept fixed)
\bea
\sum_{j=1}^{N_c}\bar\psi t^a_j\psi_i^y &=& 0\\
\sum_{j,k=1}^{N_c}\bar\psi t^a_j t^y_k \psi_i^x &=& \sqrt{\frac{2}{N_c}}
\left( \frac{i}{4}f_{axy} - \frac14 d_{axy}\right)\qquad\,\,\,\,\, (k=i\,, j\neq k)\\
&=& \sqrt{\frac{2}{N_c}}
\left( -\frac{i}{4}f_{axy} + \frac14 d_{axy}\right)\qquad (k\neq i\,, j\neq k)
\nonumber
\eea
One finds
\bea\label{quark}
I_1=0\,,\qquad
I_2 = -\frac12\alpha_s \frac{N_c^2-2}{N_c^{\frac52}}{\cal I}(p)
\sum_{i=1}^{N_c}\left({\cal M}^2_{i}(p)-{\cal M}^3_{i}(p)\right)\simeq N_c^{-\frac12}\,,
\eea
where we took into account the scaling law for the strong coupling $\alpha_s N_c=$
const. and that each of the $N_c$ terms in the sum over $i$ is of order 1.

The same scattering amplitude can be expressed as a sum of hadronic diagrams
with $B$ and $B_{\rm h}$ appearing as intermediate states in the $s$-channel.
The vertices in these graphs can be parametrized in terms of the following operators.
In addition to $X^{ia}\,,Z^{ia}$ parametrizing
pion couplings to the ground state \cite{DJM} and hybrid baryons
respectively, we introduce four new couplings defined as follows.

\begin{itemize}
\item $W^{ia}$ describes nondiagonal hybrid meson coupling between ordinary
and hybrid baryons defined such that the vertex is given by a sum of terms of the
form
\bea\label{Wdef}
-\frac{i}{f_M}t(T,\ell) \,N_c^\kappa \langle B|W^{\ell a}|B_{\rm h}\rangle\,.
\eea

\item $X^{'ia}$ describes hybrid meson coupling to ordinary baryons.

\item $Z^{'ia}$ describes hybrid meson coupling to hybrid baryons.

\item $W^{'ia}$ describes nondiagonal ordinary meson coupling between ordinary
and hybrid baryons, defined analogously to (\ref{Wdef}).
\end{itemize}

For generality, we assumed that the meson has an arbitrary spin $J$. Then
the meson coupling in a $p$-wave can be written in terms of a tensor $t(T,\ell)$
with $T=J,J\pm 1$ constructed from the meson polarization vector and its 
momentum as in (\ref{tdef}) for $J=1$. The leading $N_c$ dependence of the
vertex has been factored out, such that the $1/N_c$ expansion of $W^{i a}$ starts
with a term of order 1. The value of $\kappa$ for different mesons and hybrid baryons
can be extracted from the quark model calculations of Sect.~IV.

In terms of these operators, the total scattering amplitude for $B_{\rm h}+\pi(\vec p)
\to B+M_{\rm h}(\vec p\,')$ is written as
\bea\label{scattamp}
{\cal T} = \frac{p^i t(T,j)}{f_\pi f_M} N_c^{1+\kappa}\left\{ \frac{1}{E(\vec p)}
\left( W^{jb\dagger}Z^{ia} - X^{ia} W^{jb\dagger}\right)
+ \frac{1}{E(\vec p\,')}
\left( X^{'jb\dagger}W^{'ia} - W^{'ia} Z^{'jb\dagger}\right)\right\}\,.
\eea
For all cases considered in this paper (except for the coupling (\ref{Mares'})), 
one has $\kappa \geq 0$. Imposing consistence with the quark diagram scaling 
law (\ref{quark}) requires that the 
operator relations in (\ref{scattamp}) vanish independently to leading order
in $1/N_c$. We will be interested in the following only in the first of these
relation, which will be used to constrain the coupling $W^{jb}$
\bea\label{ccs}
& &W^{jb\dagger}_0 Z^{ia}_0 - X^{ia}_0 W^{jb\dagger}_0 = 0\\
& &X^{'jb\dagger}_0 W^{'ia}_0 - W^{'ia}_0 Z^{'jb\dagger}_0 = 0\,.
\eea

The solutions for $X^{ia}_0$ and $Z^{ia}_0$ are well known \cite{DJM}. Their 
matrix elements on tower states are given by
\bea\label{X0sol}
& &\langle J'I'\Delta';m'\alpha'|X^{ia}_0|JI\Delta; m\alpha\rangle = \\
& &\qquad g(X)\sqrt{(2I+1)(2J+1)}(-)^{2I-J+I'-\Delta+1}
\left\{ \begin{array}{ccc}
I' & 1 & I \\
J & \Delta & J' \end{array}\right\}
\delta_{\Delta \Delta'}
\langle J'm'|J1;mi\rangle \langle I'\alpha'|I1;\alpha a\rangle\nonumber
\eea
and analogous for $Z^{ia}_0$.

In the following we will solve the consistency conditions (\ref{ccs}) for the
matrix elements of $W^{ia}$. First, we parametrize the matrix elements of this
operator on tower states in terms of reduced matrix elements $W,\bar W$ as
\bea
& &\langle J'I'\Delta'(B);m'\alpha'|W^{ia}_0|JI\Delta(B_{\rm h}); m\alpha\rangle = \\
& &\qquad g(W)\sqrt{(2I+1)(2J+1)}(-)^{J+I+J'+I'}W(J'I',JI)
\langle J'm'|JT;mi\rangle \langle I'\alpha'|I1;\alpha a\rangle\nonumber\\
& &\langle J'I'\Delta'(B_{\rm h});m'\alpha'|W^{ia}_0|JI\Delta(B); m\alpha\rangle = \\
& &\qquad g(W)\sqrt{(2I+1)(2J+1)}(-)^{2J+2I}\overline{W}(J'I',JI)
\langle J'm'|JT;mi\rangle \langle I'\alpha'|I1;\alpha a\rangle\nonumber\,.
\eea
With this choice for the normalization factors, the reduced matrix elements
satisfy the symmetry relation
\beq
W(J'I',JI) = \overline{W}(JI,J'I')\,.
\eeq
It is easy now to transform the consistency conditions (\ref{ccs}) into
algebraic equations for the  reduced matrix elements, by projecting them
onto channels with well-defined spin $H$ and isospin $K$ \cite{DJM}. We obtain
\bea
& &\sum_{J_1 I_1} (2I_1+1)(2J_1+1)  (-)^{-J'+J_1}
\left\{ \begin{array}{ccc}
I' & 1 & I_1 \\
J_1 & \Delta' & J' \end{array}\right\}
\left\{ \begin{array}{ccc}
J & 1 & H \\
J' & T & J_1 \end{array}\right\}
\left\{ \begin{array}{ccc}
I & 1 & K \\
I' & 1 & I_1 \end{array}\right\}
W(J_1 I_1,JI) \\
& &\qquad\qquad =(-)^{I-2J+K+\Delta'-\Delta+T-1}
\left\{ \begin{array}{ccc}
K & 1 & I \\
J & \Delta & H \end{array}\right\}
W(J'I',HK)\,.\nonumber
\eea
The solution of this equation can be written directly in analogy to Eq.~(3.59)
in \cite{PY1} and is given by a sum of three 9$j$ symbols with undetermined 
coefficients\footnote{There is an additional arbitrariness of this solution, 
manifested
in the possible presence of a phase factor $(-)^{2Jn_1 + 2I n_2}$, with 
$n_1,n_2$ integers.}
\bea\label{ccsol}
W(J' I',JI) = \sum_{y=T,T\pm 1} c_y(\Delta',\Delta)
\left\{ \begin{array}{ccc}
\Delta' & I' & J' \\
\Delta & I & J \\
y & 1 & T \end{array}\right\}\,.
\eea

It is important to note that, while phrased in the language of hybrid baryons
couplings, the consistency conditions (\ref{ccs}) and their solution (\ref{ccsol})
have a wider validity. They give at the same time the general solution for
couplings of mesons of arbitrary spin to excited and ground state baryons
(assuming only that the large $N_c$ scaling of the vertices is such that the
consistency conditions (\ref{ccs}) still hold).
Therefore the results of this paper extend the results of \cite{PY1}
to include the couplings of mesons with any spin.

In the following we quote the results for the coefficients $c_y(\Delta',\Delta)$
in the quark model, using the results of Sec.~IV. This will illustrate the
solution (\ref{ccsol}) on a few particular cases.

For final ground state baryons containing only $u,d$ quarks $(\Delta'=0)$,
the solution (\ref{ccsol}) reduces to the following simpler expression
containing only 6$j$ symbols
\bea\label{Wdelta=0}
W(I',JI\Delta) = c_\Delta \frac{(-)^{1+J+I'+\Delta}}
{\sqrt{(2I'+1)(2\Delta+1)}}
\left\{ \begin{array}{ccc}
T & 1 & \Delta \\
I & J & I' \end{array}\right\}\,,\qquad
(\Delta = T, T\pm 1)\,.
\eea
To compare this against the results of Sec.~IV.A, the matrix elements of the
latter must be transformed  from the basis $|(IP)S,J_g;J\rangle$ to the basis
$|(PJ_g)\Delta,I;J\rangle$. The connection between the two is a simple recoupling
relation (Eq.~(3.23) in \cite{PY1})
\bea\label{Deltabasis}
|(PJ_g)\Delta,I;J\rangle = (-)^{-I-J}\sum_S\sqrt{(2S+1)(2\Delta+1)}
\left\{ \begin{array}{ccc}
I & 1 & S \\
1 & J & \Delta \end{array}\right\} 
|(IP)S,J_g;J\rangle\,.
\eea

The quark model matrix element (\ref{Mares}) for hybrid baryon decays into 
mesons with $J^{PC}=0^{-+}\,, 1^{--}$ gives, when expressed in the basis 
(\ref{Deltabasis}), 
\bea
W(I',JI\Delta) = \frac{1}{\sqrt6}N_c (-)^{J-I'+1}
\sqrt{\frac{2\Delta+1}{2I'+1}}
\left\{ \begin{array}{ccc}
I & 1 & I' \\
1 & J & \Delta \end{array}\right\} 
\langle 0|\!| gB|\!|\chi\rangle
\eea
which can be seen to have the form (\ref{Wdelta=0}) with $T=1$.
In a similar way, the matrix elements (\ref{MT}) can be transformed to the 
$|(PJ_g)\Delta,I;J\rangle$ basis with the result
\bea
W(I',JI\Delta) =  N_c (-)^{J-I'+\Delta+1}
\sqrt{\frac{(2\Delta+1)(2T+1)}{2I'+1}}
\left\{ \begin{array}{ccc}
\Delta & T & 1 \\
1 & 1 & 1 \end{array}\right\} 
\left\{ \begin{array}{ccc}
I & 1 & I' \\
T & J & \Delta \end{array}\right\} 
\langle 0|\!| gB|\!|\chi\rangle\,.
\eea
This takes also the form (\ref{Wdelta=0}).

The structure of the mixing $B-B_{\rm h}$ induced by the chromomagnetic term (\ref{mix})
takes on a more transparent form in the basis of the states 
$|(PJ_g)\Delta,I;J\rangle$. One obtains for the mixing amplitude the result
\bea\label{75}
& &\langle I'm'\alpha'|\frac{g}{2m_Q}\sum_{n=1}^{N_c}\vec\sigma_n\cdot\vec B(\vec r_n)|
(PJ_g)\Delta,I;Jm\alpha\rangle =\\
& &\qquad\qquad \frac{1}{2m_Q} \frac{N_c}{\sqrt2} (-)^{2I}\delta_{\Delta 0}
\, \langle 0|\!| gB|\!|\chi\rangle\,
\delta_{JI'}\delta_{II'}
\delta_{mm'}\delta_{\alpha\alpha'}\,,\nonumber
\eea
which shows that only the members of the $\Delta=0$ tower mix with the ground
state ordinary baryons. Therefore only the hybrid baryons  with 
symmetric orbital wavefunction (\ref{Hartree}) with $\Delta=0$ can be considered 
as being ill-behaved in the large-$N_c$ limit. The remaining towers (\ref{Delta1}) and
(\ref{Delta2}) with $\Delta=1,2$ are not affected by this anomaly. 

The matrix element (\ref{Mares'}) for the hybrid baryons with mixed symmetry orbital 
wavefunction cannot be expressed as (\ref{Wdelta=0}).
This is easily explained by noting that the $N_c$ scaling of this particular
coupling is not strong enough so that it should satisfy a consistency
condition (\ref{ccs}). Therefore the result (\ref{Mares'}) does not receive a 
model-independent justification
in the large $N_c$ limit and must be regarded as a mere consequence of the quark 
model. The same is true for the matrix element of the chromomagnetic moment 
responsible for the mixing $B-B_{\rm h}$ for the states $\Psi'_{B_{\rm h}}$ (\ref{mix'}). 
However, the coupling (\ref{MTres'}) does
grow fast enough with $N_c$ so that the consistency conditions constrain it.
We obtain for this matrix element
\bea
W(I',JI\Delta) = \frac{1}{\sqrt6}N_c (-)^{J+I'+T+1}
\sqrt{\frac{2T+1}{2I'+1}}
\left\{ \begin{array}{ccc}
1 & 1 & T \\
J & I' & I \end{array}\right\} \langle 0|\!| gB|\!|\chi\rangle
\eea
which has precisely the form (\ref{Wdelta=0}) with $\Delta=1$.

\section{Conclusions}

We have studied in this paper the existence and properties of hybrid baryons from
the point of view of the large-$N_c$ expansion. Our arguments are limited to
the case of baryons made up of heavy quarks, for which the constituent picture
is exactly valid. For this case a nonrelativistic Hartree description for quarks
of these states can be introduced, similar to the one commonly used for ordinary baryons 
in the large $N_c$ limit \cite{Wi}. Two types of hybrid baryons are discussed,
the ground state hybrids (\ref{Hartree}) and hybrids with mixed symmetric orbital
wavefunction (\ref{Hartree1}). Their spin-flavor structure
is similar to that of ordinary orbitally excited baryons and ground state baryons
respectively.
The variational equation for the
Hartree wavefunction turns out to coincide, in the large $N_c$ limit, with the
one corresponding to the ordinary baryons. This proves the existence of hybrid
baryons with heavy quarks in the large $N_c$ limit. 

Using the Hartree picture one can study static properties of the hybrid baryons
such as their mass and mixing with ordinary baryons. A surprising result
concerns the large mixing amplitude of the ground state hybrid baryons with the
ordinary baryons. A similar observation has been made in \cite{2} in the framework
of the bag model. These authors find that for $N_c=3$ the hybrid baryons mix strongly
with the ground state baryons. 
In contrast, the mixing of the hybrids with mixed symmetry orbital wave functions
(\ref{Hartree1}) is suppressed in the large $N_c$ limit. 
While the applicability of these large $N_c$ arguments to the real world is
probably rather limited, this result suggests that
the states of the $\Delta=0$ tower of hybrids might be very broad (as discussed 
in the main text) and thus difficult to observe. 

The scalings laws of the hybrid baryons' couplings can be easily determined
with the help of their Hartree wavefunctions. These are used in turn to
write down consistency conditions for the couplings, similar to those introduced
for determining the pion couplings of baryons by Dashen, Jenkins and Manohar 
\cite{DJM}. These consistency conditions are solved explicitly and their
solutions are shown to coincide with the quark model result for these
couplings. 

In principle these results could be used to distinguish between positive parity
hybrid baryons and radial excitations of the ordinary baryons (Roper resonances)
by comparing their strong decay amplitudes into ordinary baryons and mesons. 
The dependence of these two amplitudes on the quantum numbers of the respective 
states is different:
Eqs.~(\ref{Wdelta=0}) for the former and (\ref{X0sol}) for the latter respectively.
Unfortunately, at present the practical power of this method is likely to be 
rather limited. Indeed, in order to use our predictions,
each of these states has to be unambiguously assigned to one large $N_c$ tower (as the 
solutions (\ref{Wdelta=0}), (\ref{X0sol}) depend on the tower quantum number $\Delta$). 
This, in turn, would be possible only if all, or at least most of the expected 
hybrid states have been identified. 
For the negative parity baryons this assignment is relatively easy and has
been presented in \cite{PY2}. At the present moment this is far from being 
accomplished for the positive parity states. One can only hope that with improved 
experimental data, the results of this paper will be put to test in a not too
distant future.

\acknowledgements
The research of D.P. has been supported by the Ministry of Science and the Arts
of Israel. The work of C-K.C. and T.M.Y. was supported in part by the 
National Science Foundation.

\thispagestyle{plain}
\begin{figure}[hhh]
 \begin{center}
 \mbox{\epsfig{file=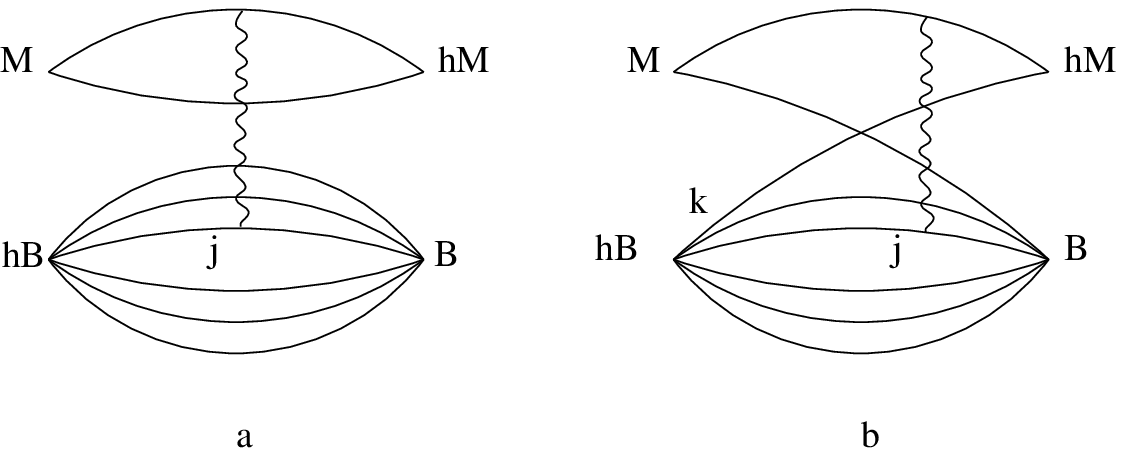,width=9cm}}
 \end{center}
 \caption{Typical diagrams showing color flows for scattering amplitudes
$B_{\rm h}+M(p)\to B+M_{\rm h}(p')$ to order $\alpha_s$. }
\label{fig2}
\end{figure}

\end{document}